\documentclass{article}
\usepackage{spconf,amsmath,graphicx,hyperref}
\usepackage{algorithm}
\usepackage{algpseudocode}
\usepackage{times}  
\usepackage{helvet}
\usepackage{courier}  
\usepackage{caption}
\usepackage{amssymb}
\usepackage{amsmath}
\usepackage{amsthm}

\usepackage{xcolor}
\usepackage{colortbl}

\usepackage{multirow}    
\usepackage{array}       
\usepackage{booktabs}
\usepackage{subcaption}
\usepackage{float}
\usepackage{newfloat}
\usepackage{listings}

\newcommand{\tool}{\textsc{VCD-Rnk}}

\title{The Cream Rises to The Top: \\Efficient Reranking Method for Verilog Code Generation}
\name{Guang Yang$^{\clubsuit}$ \quad Wei Zheng\thanks{Wei Zheng is corresponding author.}$^{\clubsuit}$ \quad Xiang Chen$^{\diamondsuit}$ \quad Yifan Sun$^{\dagger}$ \quad Fengji Zhang$^{\heartsuit}$ \quad Terry Yue Zhuo$^{\spadesuit}$}
\address{$^{\clubsuit}$ Northwest Polytechnical University, Shaanxi, China \\
$^{\diamondsuit}$Nantong University, Jiangsu, China\\
$^{\dagger}$Minzu University of China, Beijing, China\\
$^{\heartsuit}$City University of Hong Kong, Hong Kong, China\\
$^{\spadesuit}$Monash University, Victoria, Australia}

\begin{document}
%
\maketitle
\begin{abstract}
    LLMs face significant challenges in Verilog generation due to limited domain-specific knowledge. 
    While sampling techniques improve $\mathrm{pass@}k$ metrics, hardware engineers need one trustworthy solution rather than uncertain candidates.
    To bridge this gap, we formulate it as a semantic alignment problem between requirements and Verilog implementations, and propose {\tool}, a discriminator model tailored for efficient Verilog code reranking.
    Specifically, {\tool} incorporates Verilog-specific reasoning by distilling expert knowledge across three dimensions: code semantic analysis, test case generation, and functional correctness assessment.
    By explicitly simulating the above reasoning processes during inference, {\tool} effectively avoids computationally intensive test execution in existing methods.
\end{abstract}
\section{Introduction}
\label{sec:intro}

Modern integrated circuit design complexity necessitates automated Hardware Description Language (HDL) generation (e.g., Verilog, VHDL, and SystemVerilog) to reduce costs and engineering workload~\cite{flake2020verilog, alsaqer2024potential}. 
While LLMs have shown promising capabilities in software code generation~\cite{yang2024chain}, they exhibit limited performance in Verilog generation due to insufficient domain knowledge~\cite{ho2025verilogcoder, liudeeprtl}, and retraining multiple models remains prohibitively resource-intensive~\cite{shi2024continual, de2024towards}.



\begin{figure}[t]
    \centering
    \includegraphics[width=\linewidth]{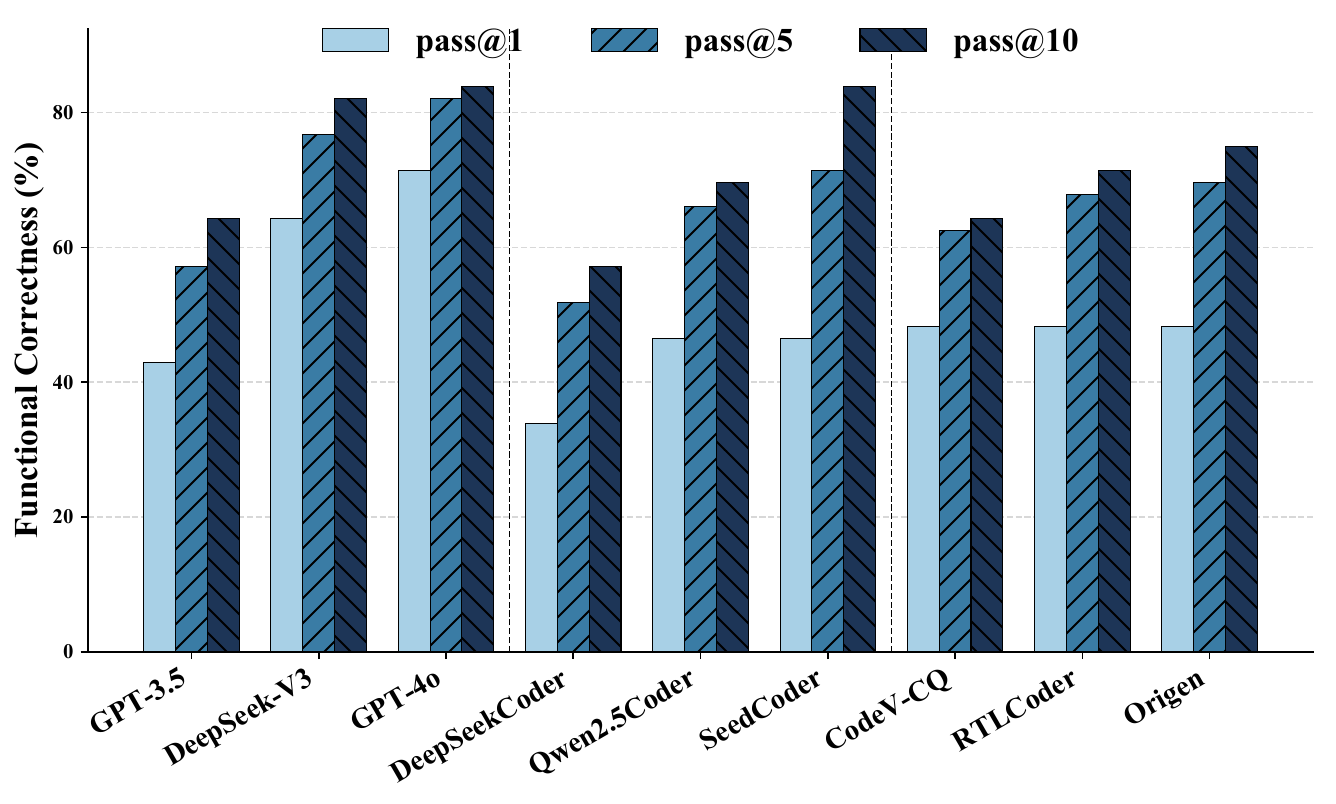}
    \caption{Performance on ResBench benchmark.}
    \label{fig:improvement_ratio}
    \vspace{-2em}
\end{figure}

Current methods~\cite{cui2024origen, liu2024rtlcoder} improve Verilog code quality through sampling techniques, generating multiple candidates and calculating $\mathrm{pass@}k$ metrics. 
This strategy leverages the observation that correct implementations often exist within a larger candidate pool, as evidenced by significantly higher $\mathrm{pass@}k$ compared to $\mathrm{pass@}1$ performance across diverse LLMs (as shown in Figure~\ref{fig:improvement_ratio}).

However, $\mathrm{pass@}k$ quantifies model potential but lacks a mechanism to identify optimal implementations. 
In practical scenarios, engineers require a single reliable solution rather than multiple uncertain candidates~\cite{chencodet}, making it challenging to select the best implementation from top-$k$ candidates.

To address these limitations, we introduce {\tool}, a lightweight discriminator model for efficient Verilog code reranking that incorporates Verilog-specific reasoning through knowledge distillation. 
{\tool} emulates human engineers' evaluation across three dimensions (code semantic analysis, test case generation, and functional correctness assessment) while eliminating test execution overhead.
Our method constructs a discrimination dataset $\langle\text{nl}, \text{verilog}, \text{label}\rangle$, employs collaborative distillation with dual teacher models to create \textbf{VerilogJudge-47K}, and fine-tunes Qwen3-4B with majority voting for candidate selection, achieving 10.4-25.8\% improvement in $\mathrm{pass@}1$ performance across multiple LLMs.



Our contributions are:
(1) We formulate Verilog code reranking as a functional consistency evaluation between natural language and code.
(2) We construct a reasoning dataset and a customized discriminator model {\tool} tailored for efficient Verilog code reranking.
(3) We achieve 10.4-25.8\% improvement in $\mathrm{pass@}1$ performance across multiple LLMs while maintaining lower computational overhead.

\section{Background and Related Work}
\label{sec:related_work}

\subsection{Code Generation for Verilog}

Verilog features concurrency, timing-dependent behavior, and hardware-specific syntax that create unique challenges for code generation systems. Given dataset $\mathcal{D} = \{(x_i,y_i)\}_{i=1}^{|\mathcal{D}|}$ where $x_i$ is natural language specification and $y_i$ is Verilog implementation, a code generation model computes $p_{\theta}(y|x) = \prod_{t=1}^{|y|}p_{\theta}(y_t|x, y_{<t})$.

\subsection{Pass@k Metric}


The $\mathrm{pass@}k$ metric assesses whether at least one correct implementation exists among $k$ candidates:
\begin{equation}
\mathrm{pass@}k := \mathbb{E}_{x \sim \mathcal{D}} \left[ 1-\frac{\binom{n-c(x)}{k}}{\binom{n}{k}} \right]
\end{equation}
where $n$ is total candidates, $c(x)$ is correct implementations for problem $x$.

\subsection{Reranking Method for Code Generation}

The $\mathrm{pass@}k$ metric quantifies model performance but lacks a mechanism to select optimal implementations from multiple candidates. 

Given $k$ candidates $\mathcal{Y}_k = \{\hat{y}_1, \ldots, \hat{y}_k\}$ for input $x$, a reranking function $R: \mathcal{X} \times \mathcal{Y} \rightarrow \mathbb{R}$ selects:
\begin{equation}
\hat{y}^* = \arg\max_{\hat{y}_i \in \mathcal{Y}_k} R(x, \hat{y}_i)
\end{equation}

Existing methods include manual selection~\cite{blocklove2023chip}, mutual information-based CodeReviewer~\cite{liu2023verilogeval}, and test-based CodeT~\cite{chencodet}. However, they perform poorly in Verilog domain and introduce significant computational overhead.

\section{Method}
\label{sec:method}

\begin{figure}[t]
    \centering
    \includegraphics[width=\linewidth]{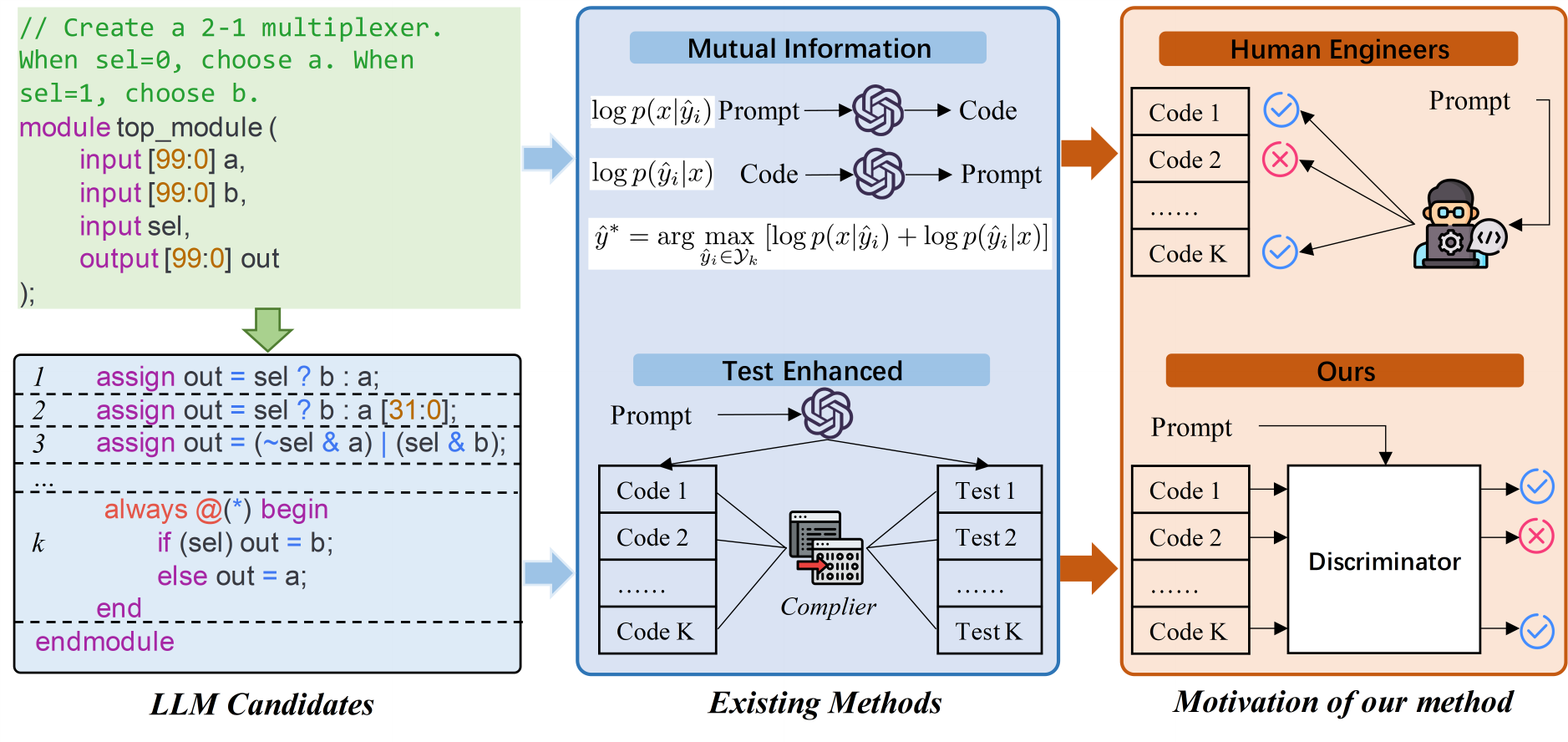}
    \caption{Comparison of existing methods and {\tool}.}
    \label{fig:compare}
    \vspace{-0.5em}
\end{figure}

As shown in Figure~\ref{fig:compare}, 
{\tool} directly learns the semantic mapping between specifications and implementations without test case execution and complex consensus calculations.


\begin{algorithm}[t!]
    \caption{Collaborative Knowledge Distillation}
    \label{alg:collaborative_distillation}
    \begin{algorithmic}[1]
    \Require Seed dataset $\mathcal{D}$; Teacher models $T_1$, $T_2$; Code Generator $G$
    \Ensure Distilled dataset $\mathcal{D}_{distill}$
    \State $\mathcal{D}_{candi} \gets \emptyset$
    \For{each $(x, y, t) \in \mathcal{D}$}
        \State $\{y^1,...,y^k\} \gets G(x)$
        \For{each $y^j$ in $\{y^1,...,y^k\}$}
            \State $label_j \gets \textsc{Execute}(y^j, t)$
            \State $\mathcal{D}_{candi}$.append$\{(x, y^j, label_j)\}$
        \EndFor
    \EndFor
    \State $\mathcal{D}_{distill} \gets \emptyset$
    \For{each $(x, y, label) \in \mathcal{D}_{candi}$}
        \If{$\textsc{$T_1$}(x, y).pred = label$}
            \State $\mathcal{D}_{distill}$.append$\{(x, y, label, \textsc{$T_1$}(x, y).reason)\}$
        \ElsIf{$\textsc{$T_2$}(x, y).pred = label$}
            \State $\mathcal{D}_{distill}$.append$\{(x, y, label, \textsc{$T_2$}(x, y).reason)\}$
        \EndIf
    \EndFor
\end{algorithmic}
\end{algorithm}

\begin{figure*}[t]
    \centering
    \includegraphics[width=0.92\linewidth]{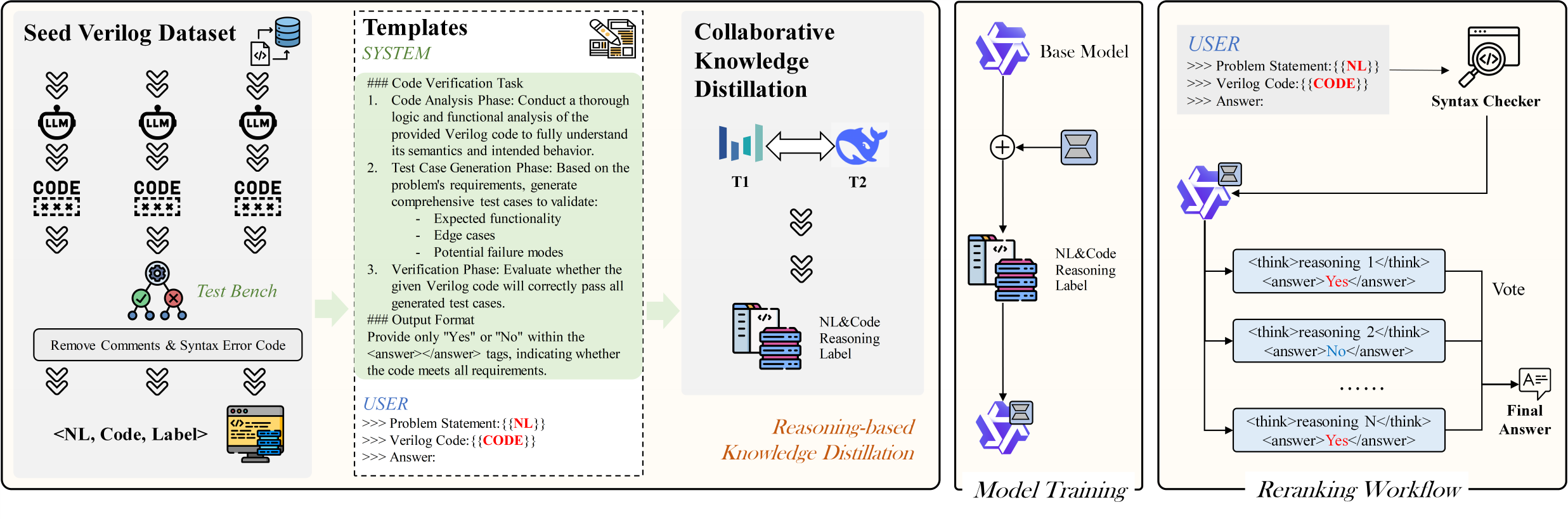}
    \caption{Overview of our method.}
    \label{fig:approach}
    \vspace{-0.5em}
\end{figure*}

\begin{figure*}[t!]
    \centering
    \includegraphics[width=0.92\textwidth]{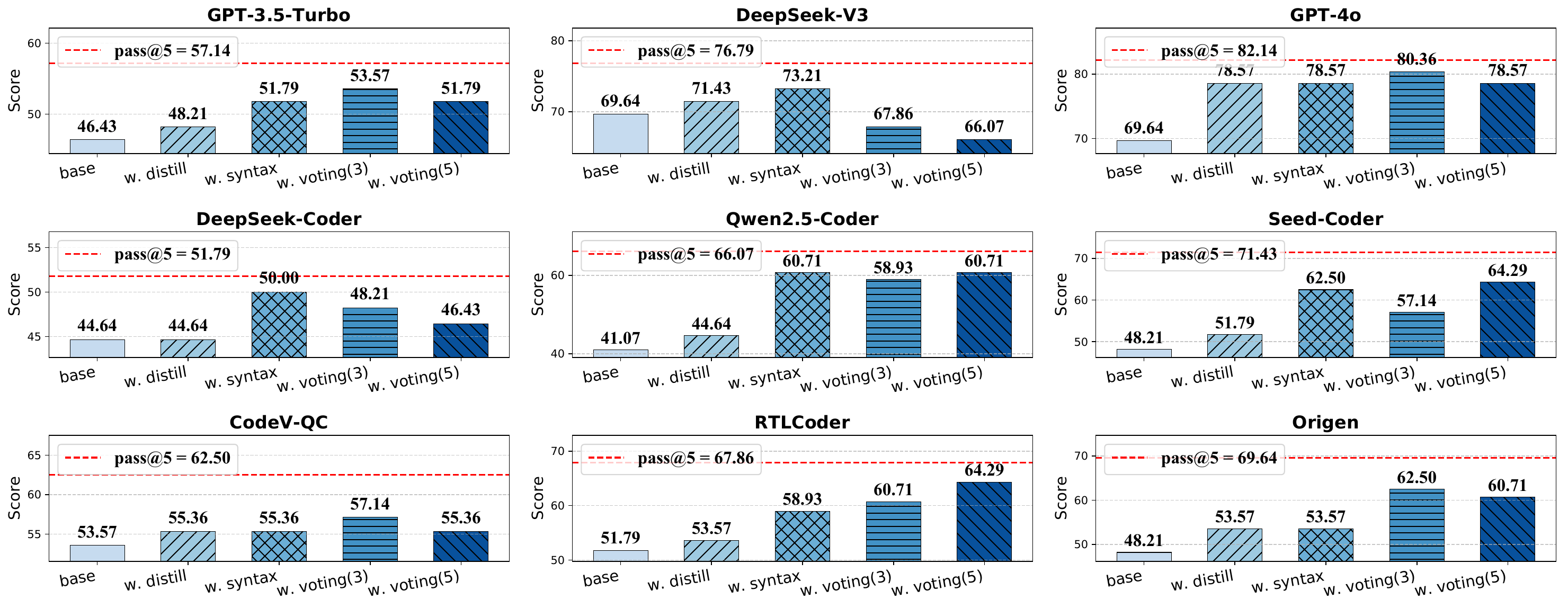}
    \caption{Ablation results on ResBench.}
    \label{fig:ablation_resbench}
    \label{fig:ablation}
    \vspace{-1.5em}
\end{figure*}

\subsection{Problem Formulation}
Formally, given a natural language specification $x \in \mathcal{X}$ and $k$ candidate implementations $\mathcal{Y}k = \{\hat{y}_1, \hat{y}_2, \ldots, \hat{y}_k\} \subset \mathcal{Y}$ generated by a model $p{\theta}(\cdot|x)$, we define a discriminator $F_{\phi}: \mathcal{X} \times \mathcal{Y} \rightarrow [0,1]$ parameterized by $\phi$ that computes the probability of functional correctness:
\begin{equation}
F_{\phi}(x, \hat{y}i) = p(\text{correct}|x, \hat{y}_i; \phi)
\end{equation}
The discriminator is trained to maximize the likelihood of correct classification:
\begin{equation}
\begin{aligned}
\mathcal{L}{\text{disc}}(\phi) = & -\mathbb{E}{(x,y) \sim \mathcal{D}} [c(y) \log F_{\phi}(x,y) + \\
& (1-c(y)) \log(1-F_{\phi}(x,y)) ]
\end{aligned}
\end{equation}

During inference, the optimal implementation is selected by:
\begin{equation}
\hat{y}^* = \arg\max_{\hat{y}i \in \mathcal{Y}_k} F_{\phi}(x, \hat{y}i)
\end{equation}

\subsection{{\tool} Design}
Figure~\ref{fig:approach} shows {\tool} design.
{\tool} extracts Verilog-specific knowledge through collaborative dual-teacher distillation using VeriCoder's dataset~\cite{wei2025vericoder}, trains a discriminator to learn semantic mapping between specifications and implementations, then reranks candidates based on discriminator decisions.


\label{sec:discriminator}
\noindent\textbf{(1) Collaborative Knowledge Distillation}
Algorithm~\ref{alg:collaborative_distillation} outlines our dual-teacher distillation approach. 
We use Seed-Coder~\cite{zhang2025seed} to sample 10 implementations per specification. 
For knowledge distillation, we employ doubao-seed-1.6 as primary teacher ($T_1$) and DeepSeek-R1-671B as secondary teacher ($T_2$) to handle cases where $T_1$ fails, creating \textbf{VerilogJudge-47K} dataset.




\noindent\textbf{(2) Model Training}
We fine-tune Qwen3 on \textbf{VerilogJudge-47K} using LoRA~\cite{hu2022lora}, which freezes pre-trained weights $W_0 \in \mathbb{R}^{d \times k}$ while introducing trainable matrices $A \in \mathbb{R}^{r \times k}$ and $B \in \mathbb{R}^{d \times r}$ ($r \ll \min(d,k)$). Weight updates follow $W = W_0 + BA$, reducing parameters from $dk$ to $r(d + k)$.

\noindent\textbf{(3) Reranking Workflow}
\label{sec:reranking_workflow}
Our workflow uses syntax checker for filtering, then majority voting for final selection. We perform multiple inference passes and calculate:
\begin{equation}
F_{\phi}(x, \hat{y}_i) = \frac{\sum_{j=1}^{m} \mathbb{I}[F_{\phi}^{(j)}(x, \hat{y}_i) = 1]}{m}
\end{equation}
where $F_{\phi}^{(j)}(x, \hat{y}_i)$ represents the prediction from the $j$-th inference pass.




\section{Experiments}
\label{sec:experiments}

\subsection{Setup}

\noindent\textbf{Datasets} We evaluate on two diverse real-world Verilog benchmarks: \textbf{RTLLM-v1.1}~\cite{lu2024rtllm} and \textbf{ResBench}~\cite{guo2025resbench}.


\noindent\textbf{Evaluation Metrics}
We follow standard practice in code generation literature by using $\mathrm{pass@}k$,
and $\mathrm{pass@}1$ after applying different reranking strategies to the top-$k$ candidates.

\noindent\textbf{Studied Models}
We evaluate across general-purpose (GPT-3.5-Turbo, GPT-4o, DeepSeek-V3), code-specialized (DeepSeek-Coder, Qwen2.5-Coder, Seed-Coder), and Verilog-specialized (CodeV-QC~\cite{zhao2024codev}, RTLCoder~\cite{liu2024rtlcoder}, Origen~\cite{cui2024origen}) LLMs.

\noindent\textbf{Baselines}
We compare against Probability, CodeReviewer, CodeRank, and CodeT methods. 
We use CodeRankEmbed\footnote{\url{https://github.com/cornstack/CodeRankEmbed}} for CodeRank and generate five test cases per implementation for CodeT.


\begin{table*}[t!]
    \centering
    \resizebox{0.92\textwidth}{!}{
    \begin{tabular}{l cccccc|>{\itshape}c >{\itshape}c |c}
    \toprule
    \multirow{2}{*}{\textbf{Model}} & \multirow{2}{*}{\textbf{Pass@1}} & \multicolumn{7}{c}{\textbf{Pass@1 (Reranking k=5)}} & \multirow{2}{*}{\textbf{Pass@5}} \\
 \cmidrule(lr){3-9}
 & & \textbf{Prob.} & \textbf{CodeReviewer} & \textbf{CodeRank} & \textbf{CodeT} & \textbf{Ours} & \textbf{$T_1$} & \textbf{$T_2$}\\
    \midrule
    \midrule
    \multicolumn{10}{c}{\textbf{Evaluation on RTLLM-v1.1}} \\
    \midrule
    GPT-3.5-Turbo & 32.14 & - & - & 32.14 & \textbf{35.71} & \textbf{35.71} & 39.29 & 35.71 & 39.29\\
    DeepSeek-V3 & 57.14 & - & - & 57.14 & 64.29 & \textbf{67.86} & 57.14 & 57.14 & 67.86\\
    GPT-4o & 60.71 & - & - & 57.14 & 64.29 & \textbf{67.86} & 64.29 & 67.86 & 75.00\\
    \midrule
    DeepSeek-Coder & 32.14 & 28.57 & 21.43 & 28.57 & 28.57 & \textbf{32.14} & 39.29 & 28.57 & 42.86\\
    Qwen2.5-Coder & 42.86 & 39.29 & 35.71 & 39.29 & 46.43 & \textbf{50.00} & 42.86 & 46.43 & 53.57\\
    Seed-Coder & 42.86 & 39.29 & 50.00 & 46.43 & 42.86 & \textbf{53.57} & 53.57 & 39.29 & 64.29\\
    \midrule
    CodeV-QC & 46.43 & 46.43 & 42.86 & 42.86 & 46.43 & \textbf{53.57} & 53.57 & 53.57 & 60.71\\
    RTLCoder & 42.86 & 28.57 & 32.14 & 28.57 & \textbf{42.86} & 39.29 & 35.71 & 39.29 & 50.00\\
    Origen & 53.57 & 53.57 & \textbf{57.14} & 46.43 & \textbf{57.14} & 53.57 & 57.14 & 50.00 & 71.43\\
    \midrule
    Avg. & 45.64 & 39.29 & 39.88 & 42.06 & 47.62 & \textbf{50.40} & 49.21 & 46.43 & 58.33\\
    \midrule
    \midrule
    \multicolumn{10}{c}{\textbf{Evaluation on ResBench}} \\
    \midrule
    GPT-3.5-Turbo & 42.86 & - & - & 42.86 & 46.43 & \textbf{53.57} & 51.79 & 48.21 & 57.14\\
    DeepSeek-V3 & 64.29 & - & - & 64.29 & 71.43 & \textbf{73.21} & 69.64 & 64.29 & 76.79\\
    GPT-4o & 71.43 & - & - & 66.07 & 71.43 & \textbf{80.36} & 76.79 & 75.00 & 82.14\\
    \midrule
    DeepSeek-Coder & 33.93 & 42.86 & 25.00 & 26.79 & 46.43 & \textbf{50.00} & 50.00 & 39.29 & 51.79\\
    Qwen2.5-Coder & 46.43 & 39.29 & 53.57 & 39.29 & 46.43 & \textbf{60.71} & 60.71 & 60.71 & 66.07\\
    Seed-Coder & 46.43 & 44.64 & 41.07 & 46.43 & 53.57 & \textbf{64.29} & 60.71 & 55.36 & 71.43\\
    \midrule
    CodeV-QC & 48.21 & 50.00 & 46.43 & 48.21 & 50.00 & \textbf{57.14} & 55.36 & 57.14 & 62.50\\
    RTLCoder & 48.21 & 46.43 & 35.71 & 37.50 & 51.79 & \textbf{64.29} & 58.93 & 53.57 & 67.86\\
    Origen & 48.21 & 41.07 & 44.64 & 48.21 & 50.00 & \textbf{62.50} & 57.14 & 48.21 & 69.64\\
    \midrule
    Avg. & 50.00 & 44.05 & 41.07 & 46.63 & 54.12 & \textbf{62.90} & 60.12 & 55.75 & 67.26\\
    \bottomrule
    \end{tabular}
    }
    \caption{Performance comparison (\%) of various model categories across two Verilog benchmarks.}
    \label{tab:model-comparison}
    \vspace{-0.5em}
\end{table*}

\subsection{Main Results}

The main results are presented in Table \ref{tab:model-comparison}. For closed-source models, probability information is unavailable, resulting in ``-" for Probability and CodeReviewer methods.


\textbf{(1) Comparison with Baselines} 
{\tool} achieves superior performance on RTLLM-v1.1 (50.40\%) and ResBench (62.90\%), with improvements of +5.8-16.2\% over CodeT and +10.4-25.8\% over original Pass@1, demonstrating better semantic alignment capture than existing methods.



\textbf{(2) Comparison with Teacher Models} 
{\tool} successfully distills complementary reasoning from both teachers, outperforming individual models, showing effective dual-teacher distillation.


\textbf{(3) Comparison with Theoretical Upper Bound} 
{\tool} achieves 86.4\% (50.40/58.33) and 93.5\% (62.90/67.26) of theoretical upper bounds on RTLLM-v1.1 and ResBench respectively, substantially narrowing the performance gap.


\textbf{(4) Statistical Significance} 
To rigorously validate the improvements observed in $\mathrm{pass@}1$ metrics, we conducted Wilcoxon signed-rank tests comparing {\tool} against baselines, showing {\tool}'s improvements are statistically significant ($p < 0.01$) in 14/18 model configurations.

\subsection{Ablation Results}
To evaluate individual component contributions, we conducted ablation studies on {\tool}'s three main components in Figure \ref{fig:ablation}.
\textbf{Knowledge Distillation} provides substantial improvement (+9.52\% on average), confirming Verilog-specific knowledge enhances specification-implementation consistency evaluation.
\textbf{Syntax Checker} and \textbf{Majority Voting} both offer consistent improvement, demonstrating the importance of filtering syntactically incorrect candidates.

\subsection{Efficiency Comparison}

We compare computational efficiency between {\tool} and CodeT across different scenarios.
CodeT requires three sequential stages: code generation (~1-5s), test generation (~1.5-5s), and test execution (~1s per iteration). 
In contrast, {\tool} achieves ~1.5s inference latency per instance, remaining substantially lower even with voting iterations. 
Importantly, {\tool} eliminates the test execution phase, providing significant deployment advantages.


\section{Conclusion}

We introduced {\tool}, a discriminative reranking method for Verilog code generation that addresses the gap between model capability and practical hardware design requirements. 
Future directions include extending our framework to other hardware description languages to enable more complex hardware design tasks.

\vfill\pagebreak


\bibliographystyle{IEEEbib}
\bibliography{refs}

\end{document}